\documentclass[a4paper]{ieeeconf}      % Use this line for a4 paper

\IEEEoverridecommandlockouts                              % This command is only needed if 
\usepackage{graphics} % for pdf, bitmapped graphics files
\usepackage{epsfig} % for postscript graphics files
\usepackage{txfonts} % this is for the \deltaup

\usepackage{mathptmx}

\usepackage{times} % assumes new font selection scheme installed

\usepackage{amsmath} % assumes amsmath package installed
\usepackage{amssymb}  % assumes amsmath package installed
\usepackage{dsfont}
\usepackage{enumerate}
\usepackage[tight]{subfigure}
\usepackage{multirow}
\usepackage{bm}

\usepackage{color}

\title{%\LARGE \bf
Graphon-based sensitivity analysis of SIS epidemics*%
}

\author{Renato Vizuete, Paolo Frasca, and Federica Garin%$^{2}$ <-this % stops a space
\thanks{*This work has been partially supported by the LabEx PERSYVAL-Lab (ANR-11-LABX-0025-01) funded by the French program Investissement d'Avenir and by ANR grant HANDY (ANR-18-CE40-0010).}
\thanks{The authors are with Univ.\ Grenoble Alpes, CNRS, Inria, Grenoble INP, GIPSA-lab, F-38000 Grenoble, France. R.~Vizuete is also with L2S, CentraleSup\'{e}lec, Paris Saclay, 3 rue Joliot Curie, 91190, Gif sur Yvette, France.
{\tt\small renato.vizuete@gipsa-lab.fr},
{\tt\small paolo.frasca@gipsa-lab.fr},\protect\\
{\tt\small federica.garin@inria.fr}.}% %end thanks
} %end author
\newcommand{\vertiii}[1]{{\left\vert\kern-0.25ex\left\vert\kern-0.25ex\left\vert #1 
    \right\vert\kern-0.25ex\right\vert\kern-0.25ex\right\vert}}

\newtheorem{definition}{Definition}
\newtheorem{theorem}{Theorem}

\newtheorem{prop}{Proposition}
\newtheorem{lemma}{Lemma}
\newtheorem{remark}{Remark}

\newcommand{\blue}[1]{{#1}}

\begin{document}

\maketitle
\thispagestyle{empty}
%\pagestyle{empty}

%%%%%%%%%%%%%%%%%%%%%%%%%%%%%%%%%%%%%%%%%%%%%%%%%%%%%%%%%%%%%%%%%%%%%%%%%%%%%%%%
\begin{abstract}
In this work, we use the spectral properties of graphons to study stability and sensitivity to noise of deterministic SIS epidemics over large networks. %using the characteristics of graphons as a tool for their estimation. Mainly, 
%The spectrum of the adjacency matrix is studied with a focus on applications concerning all the eigenvalues, which are hard to obtain for large networks. 
We consider the presence of additive noise in a linearized SIS model and  we derive a noise index to quantify the deviation from the disease-free state due to noise. For finite networks, we show that the index depends on the adjacency eigenvalues of its graph.  
We then assume that the graph is a random sample from a piecewise Lipschitz graphon with finite rank and, using the eigenvalues of the associated graphon operator, we find an approximation of the index that is tight when the network size goes to infinity. A numerical example is included to illustrate the results.
\end{abstract}

%%%%%%%%%%%%%%%%%%%%%%%%%%%%%%%%%%%%%%%%%%%%%%%%%%%%%%%%%%%%%%%%%%%%%%%%%%%%%%%%
\section{Introduction}

In recent years, the attention to the analysis of networks has increased in the scientific community due to the continuous evolution of the world towards a networked environment, where ever more connections are established at every instant, thereby generating networks with a large number of components, which we will refer to as \textit{large networks}. Traditionally, researchers have used concepts of Graph Theory for the analysis of networks, where a complete knowledge of the network graph is required for most applications. This assumption is reasonable for systems with a relatively small number of agents, but in the case of large networks, significant problems arise. Firstly, a complete and updated representation of the network may not be available because of the presence of noise and errors in data and the constant evolution of links and nodes. Secondly, even when it is possible to obtain a good knowledge of network topology, their sheer size prevents the full simulation or analysis of the dynamics, or the computation of relevant network properties, because of limitations in computational resources.
 
\blue{One of the most promising tools  to address these problems are {\em graph functions}, also called \textit{graphons}, which are limits of sequences of dense graphs \cite{lovasz2006limits,borgs2008convergent,borgs2012convergent,avella2018centrality}. %The motivation for the study of sequences of this type were real interactions, where a network increases by the addition of a node to the existing structure, with edges following a specific pattern. Furthermore, it was proved that some properties of the adjacency matrix of networks with a pattern similar to a graphon can be inferred from the latter. 
Researchers have already developed numerous applications of graphons to the analysis of network structures, including the approximation of
centrality measures \cite{avella2018centrality} and link prediction problems \cite{zhang2017estimating}. Very recently, researchers are also beginning to use graphons to study {\em dynamics on large networks}:  questions of interest include modeling power networks dynamics \cite{kuehn2019power} and epidemics \cite{gao2019spectral}, developing control methods \cite{gao2017control,gao2019optimal,gao2019graphon}, 
and studying large population games \cite{caines2018graphon,parise2019graphon}.} 

In this letter, we focus on the deterministic Susceptible-Infected-Susceptible (SIS) epidemic model, which describes a disease that can infect agents irrespective of whether they were infected earlier. This model is often interpreted as a meta-population model, where the state of each node is the fraction of infected individuals in a sub-population \cite{lajmanovich1976deterministic,bailey1986macro,pastor2015epidemic,pare2018analysis}. %fall2007epidemiological
% \todo{paragraph on sensitivity of epidemics}
% \blue{\cite{pare2017epidemic}}

Even if the analysis and control of epidemics are well studied topics, the applicability of the theoretical results is often limited by restrictive assumptions that require complete knowledge of the dynamical laws of the nodes, of their states, and of the structure of the network 
\cite{nowzari2016analysis,mei2017dynamics}.
The uncertainty in the network knowledge has been among the motivations to study epidemics by mean-field models \cite{zhu2012spreading,pastor2015epidemic}. In this paper, we take a different approach and account for uncertainties by modeling the network as a graphon.

%However, in reality this information is not accurate. 
%\todo{should we interpret our index as robustness to modeling defects or as reactivity to input? maybe better to stress the latter (which now sounds the best one to me)? Possibly we might move the following sentences to the previous paragraph (under construction) and leave this paragraph to only talk about graphons} 
%and it is usual to consider a certain level of uncertainty in the models \cite{zager2009epidemic}
The inclusion of additive noise is also frequently used to include un-modeled phenomena and features in epidemics models \cite{forgoston2011maximal,pare2017epidemic,krause2018stochastic}.
In network dynamics, the properties of robustness to noise can often be expressed through the spectral properties of the network \cite{fagnani2018introduction,ma2015mean,lovisari2012performance,garin2010survey}. Therefore, it becomes natural to look at the spectral properties of graphons to evaluate the robusteness properties of large networks described by graphons~\cite{gao2019spectral}.

Special cases of graphons are those corresponding to stochastic block models \cite{holland1983stochastic}, which are used to model the community structures that are frequent in real social networks \cite{karrer2011stochastic}. For instance, if we consider the case of spreading of epidemics in meta-populations, a natural approach is to consider each node as a small population, like a village or a neighborhood, and each block as a region or city. % where the interactions are more homogeneous.

The aim of this work is to leverage the properties of piecewise Lipschitz graphons with finite rank (that encompass stochastic block models) for the stability and sensitivity analysis of SIS epidemics over large networks. Our main contribution is to show that the spectral properties of the graphon allow to approximately evaluate stability and robustness to noise. %More specifically, the robustness to noise is  the effect of additive noise through an index that depends on the spectrum of the adjacency matrix.

%The remaining of the paper is structured as follows. 
%\paragraph*{Outline} 
In order to derive our approximation results, we introduce graphons and their relevant properties (see Section~2). We then develop the analysis of SIS epidemics over a network sampled from a graphon (Section~3): we define a suitable sensitivity index, we express it by using the spectral properties of the graph, and we approximate it by using the spectral properties of the graphon. Finally, we illustrate our results by simulations on a stochastic block model (Section~4) and comment about our results and future work (Section~5).

\section{Graphons}
This section contains the definition of graphon and related facts that will be needed in the following sections.
\subsection{Graphons: Basic Notions and Examples}
A graph is defined as a pair $G=(V,E)$ where $V\neq \{ \}$ is a finite set of vertices or nodes and $E\subseteq \{(i,j)\in V \times V:i\neq j\}$ is the set of edges. In this work, we consider simple graphs, such that they are undirected (i.e., edges with no direction), unweighted (i.e., edges without weights) and do not contain self-loops or multiedges.

The adjacency matrix of a graph $A=[a_{ij}]\in \mathbb{R}^{N\times N}$ is defined by $a_{ij}=1$ if $(i,j)\in E$ and $a_{ij}=0$ otherwise. This matrix is real symmetric non-negative with real eigenvalues ordered as $\lambda_1(A) \geq \lambda_2(A) \geq \cdots \geq \lambda_N(A)$. 

We denote by $\mathcal{W}$ the space of all bounded symmetric measurable functions $W:[0,1]^2\rightarrow \mathbb{R}$. The elements of this space are called \textit{kernels} given their connection to integral operators. The set of all kernels $W \in \mathcal{W}$ such that $0 \leq W \leq 1$ is denoted by $\mathcal{W}_0$ and its elements are called \textit{graphons}, whose name is a contraction of graph-function. By analogy with degrees in finite graphs, the {\em degree} function of a graphon is defined as
$
d_W(x):=\int_0^1W(x,y)\, dy .
$
In order to consider differences between graphons, we shall sometimes work in the set $\mathcal{W}_1$ of kernels $W$ such that $-1\leq W \leq 1$.

Every function $W \in \mathcal{W}$ defines an integral operator $T_W : L^2[0,1]\rightarrow L^2[0,1]$ by:
$$
\left(T_Wf\right)(x):=\int_0^1W(x,y)f(y)\, dy .
$$
This operator is compact and has a discrete spectrum with 0 as the only accumulation point. % such that $\lambda_N(T_W)\rightarrow 0$. 
Every nonzero eigenvalue has finite multiplicity \cite{lovasz2012large}.
A graphon $W$ is said to have finite rank if the spectrum of the associated operator contains a finite number of nonzero eigenvalues~\cite{lovasz2012large}.

A {\em step graphon} is a graphon defined as a step function. A function is called a \textit{step function} if there is a partition $S_1\cup\cdots\cup S_k$ of $[0,1]$ into measurable sets such that $W$ is constant on every product set $S_i\times S_j$ where the sets $S_i$ are the steps of $W$. This type of graphon is also called \textit{stochastic block model graphon} because of its relation to stochastic block models \cite{airoldi2013stochastic}. 
%generalize the popular Erd\H{o}s-R\'{e}nyi random graph to more species of vertices \cite{airoldi2013stochastic}
Step graphons are finite rank graphons with a rank at most equal to the number of steps. Also, graphons expressed as a finite sum of products of integrable functions have finite rank \cite{avella2018centrality}.

Each graph $G$ has an associated step graphon $W_G$ obtained by considering a uniform partition of $[0,1]$ into the intervals $B_i^N$, where $B_i^N=[(i-1)/N,i/N)$ for $i=1,\ldots, N-1$ and $B_N^N=[(N-1)/N,1]$ such that:
$$
W_{G}(x,y):=\sum_{i=1}^N\sum_{j=1}^N a_{ij}\mathds{1}_{B_i^N}(x)\mathds{1}_{B_j^N}(y) ,
$$
where $\mathds{1}_A(x)$ is the indicator function. The operator associated to the step graphon is
$$
(T_{W_G}f)(x):=\sum_{j=1}^N a_{ij}\int_{B_j^N}f(y)\, dy \;\;\text{for any} \; x\in B_i^N 
$$
and the spectrum of $T_{W_G}$ consists of the normalized spectrum of the graph (i.e., $\lambda_i(T_{W_G})=\lambda_i(A)/N$), together with infinitely many zeros.

\begin{figure}%[h]
\centering
\includegraphics[width=8.5cm]{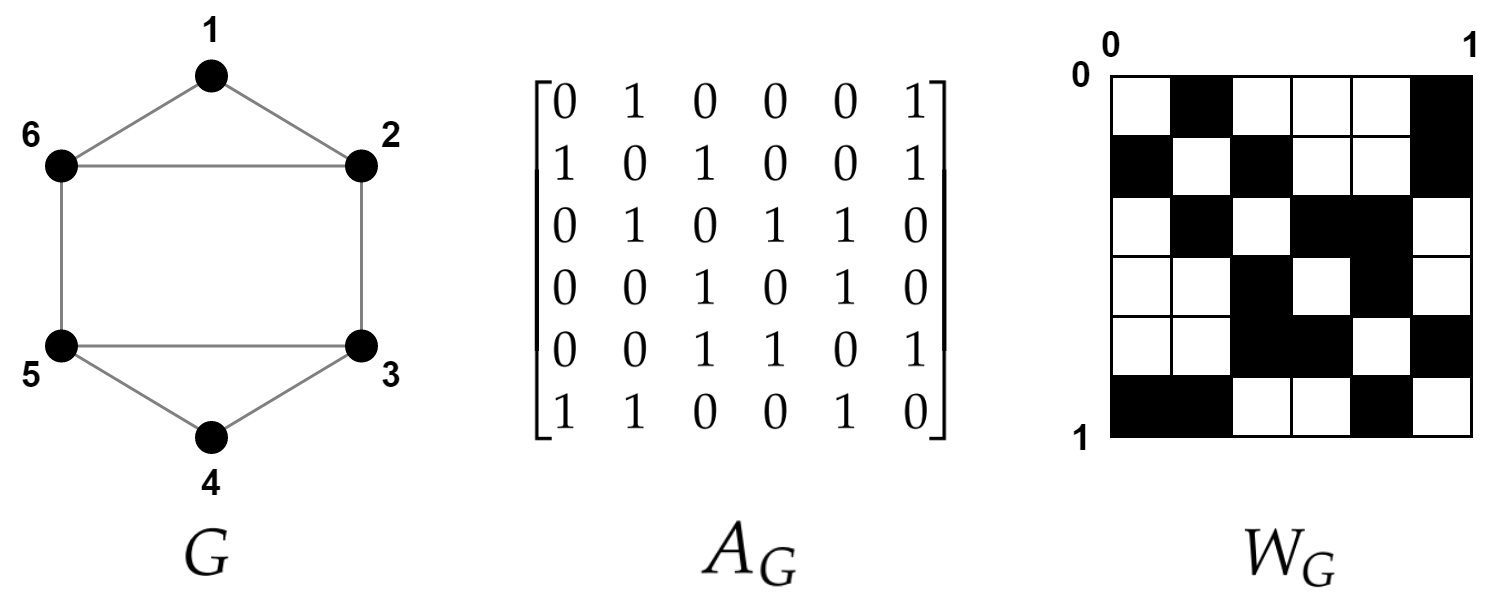}
\caption{Graph $G$, adjacency matrix $A$ and step graphon $W_G$.}
\label{f_pixeldiagram}
\end{figure}

%---------OLD
% The visualization of a graphon is usually performed with the help of a pixel picture.
% For a step graphon associated to a graph $G$, we visualize a 0 as a small white square and a 1 as a small black square as we can appreciate in Fig. \ref{f_pixeldiagram}.
%---NEW
A graphon is usually visualized with a pixel picture,
where each point $(x,y) \in [0,1]^2$ is colored with a grey level representing $W(x,y)$.
For a step graphon associated to a graph $G$, we visualize a 0 as a small white square and a 1 as a small black square as we can appreciate in Fig. \ref{f_pixeldiagram}.
%---end new

\subsection{Norms}
In the study of kernels, various norms are relevant to consider
\cite{lovasz2012large, janson2013graphons, avella2018centrality}.
For $1\leq p <\infty$, we define the $L^p$ norm of a kernel as
$$
\Vert W \Vert_{L^p}:=\left(\int_{[0,1]^2}\vert W(x,y)\vert^p dx\; dy\right)^{1/p}
$$
and its \textit{cut norm} by
$$
\Vert W \Vert_\square:=\sup_{S,T\subseteq[0,1]}\left\vert\int_{S\times T}W(x,y)\, dx\; dy\right\vert .
$$
For $W\in\mathcal{W}_1$, we have the following inequalities between $L^p$ norms and the cut norm:
% $$
% \Vert W \Vert_\square\leq\Vert W \Vert_{L^1}\leq\Vert W \Vert_{L^2}\leq 1, \qquad \Vert W \Vert_{L^2}\leq \Vert W \Vert_{L^1}^{1/2}.
% $$
$$
\Vert W \Vert_\square \leq \Vert W \Vert_{L^1}\leq
\Vert W \Vert_{L^2} \leq \Vert W \Vert_{L^1}^{1/2} \leq 1.
$$
%Additionally, $\Vert W \Vert_{L^2}\leq \Vert W \Vert_{L^1}^{1/2}$, such that these norms define the same topology.
By considering the operator $T_W$ associated to a kernel $W \in \mathcal W$, we can define the operator norm:
$$
\vertiii{T_W}:=\sup_{\substack{f \in L^2[0,1] \\ \Vert f \Vert_{L^2}=1 }}\Vert T_Wf\Vert_{L^2} .
$$
For graphons, the operator norm is equal to the largest eigenvalue of the operator: $\vertiii{T_W} = \lambda_1(T_W)$.
For the elements of $\mathcal{W}_1$, the cut and operator norms are related by:
$$
\Vert W \Vert_\square\leq\vertiii{T_W}\leq\sqrt{8}\Vert W \Vert_\square^{1/2}.
$$
Finally, we can define the Hilbert-Schmidt norm of the operator as:
$$
%\Vert T_W \Vert_{\mathrm{HS}}:=\left(\sum_i \vert \lambda_i(T_W) \vert^2 \right)^{1/2} ,
\Vert T_W \Vert_{\mathrm{HS}}^2:=\sum_i \vert \lambda_i(T_W) \vert^2. $$
For all $W \in \mathcal W$, $\Vert T_W \Vert_{\mathrm{HS}}$ is finite (i.e., kernel operators are Hilbert-Schmidt operators), and moreover
$
\Vert T_W \Vert_{\mathrm{HS}}=\Vert W \Vert_{L^2} .
$

%\mbox{}

%\begin{definition}[Random-free step graphon \cite{janson2013graphons}]
%A step graphon $W$ is said to be random-free if it has values in $\{0,1\}$. 
%\end{definition}

\subsection{Sampling and Approximation}

A graphon $W$ can be used to generate random graphs using a sampling method \cite{lovasz2012large}.
\begin{definition}[Sampled Graph \cite{avella2018centrality}]\label{samp2}
Given a graphon $W$ and a size $N \in \mathbb{N}$, we say that the graph $G$ is sampled from $W$ if it is obtained through:
\begin{enumerate}
\item Fixing deterministic latent variables $\lbrace u_i=\frac{i}{N}\rbrace _{i=1}^N$.
\item %Define the adjacency matrix $A \in \{0,1\}^{N\times N}$ of $G$ obtained by 
Taking $N$ vertices 
%$i\in \{1,\ldots ,N\}$ 
$ \{1,\ldots ,N\}$ 
and randomly adding undirected edges between vertices $i$ and $j$ independently with probability $W(u_i,u_j)$ for all $i>j$.
\end{enumerate}
\end{definition}

\begin{definition}[Piecewise Lipschitz graphon \cite{avella2018centrality}]\label{def:piecewise}
Graphon $W$ is said to be \textit{piecewise Lipschitz} if there exists a constant $L$ and a sequence of non-overlapping intervals $I_k=[\alpha_{k-1},\alpha_k)$ defined by $0=\alpha_0 < \cdots <\alpha_{K+1}=1$, for a finite non-negative integer $K$,
%\in \mathbb{N}$
such that for any $k, l$, any set $I_{kl}=I_k \times I_l$ and pairs $(x_1,y_1) \textrm{ and } (x_2,y_2)\in I_{kl}$ we have that:  \vspace{-2mm}
$$
\vert W(x_1,y_1)-W(x_2,y_2)\vert \leq L(\vert x_1-x_2 \vert+\vert y_1-y_2 \vert) . 
$$
\end{definition}\smallskip

\begin{definition}[Large enough $N$ \cite{avella2018centrality}]\label{largeN}
Given a piecewise Lipschitz graphon $W$ (as per Definition~\ref{def:piecewise}) and $\nu<e^{-1}$, $N$ is \textit{large enough} if $N$ satisfies the following conditions:
\begin{subequations}
\begin{align}
\dfrac{2}{N}<\min_{k\in \{1,\ldots , K+1\}}(\alpha_k-\alpha_{k-1}) , 
\\
\dfrac{1}{N}\log\left(\dfrac{2N}{\nu}\right)+\dfrac{1}{N}(2K+3L)<\max_{x}d_W(x) , \label{eq_condi2}\\
Ne^{-N/5}<\nu . \label{eq_condi3}
\end{align}
\end{subequations}
%
% \begin{equation*}
% \dfrac{2}{N}<\min_{k\in \{1,\ldots , K+1\}}(\alpha_k-\alpha_{k-1}) , %\label{eq_condi1}
% \end{equation*}
% %
% \begin{equation*}
% \dfrac{1}{N}\log\left(\dfrac{2N}{\nu}\right)+\dfrac{1}{N}(2K+3L)<\max_{x}d_W(x) , %\label{eq_condi2}
% \end{equation*}
% %
% \begin{equation*}
% Ne^{-N/5}<\nu . %\label{eq_condi3}
% \end{equation*}
% %
\end{definition}

\smallskip
\begin{lemma}[Theorem 1 \cite{avella2018centrality}]\label{lemma1}
Let $W$ be a piecewise Lipschitz graphon (as per Definition~\ref{def:piecewise}) and $G$ a graph with $N$ nodes sampled from $W$. Then for $N$ large enough with probability at least $1-\nu$:
$$
\vertiii{T_{W_G}-T_W}\leq \sqrt{\dfrac{4\log (2N/\nu)}{N}}+\dfrac{2\sqrt{L^2-K^2+KN}}{N}=:\phi(N) .
$$
\end{lemma}\smallskip

By considering a constant value of $\nu$, the difference between the graphons in the operator norm is bounded by: $\vertiii{T_{W_G}-T_W}=O\left((\log N/N)^{1/2}\right)$.

\smallskip
\begin{lemma}\label{lem_bnorm}
Let $W$ be a piecewise Lipschitz graphon and $G$ a graph with $N$ nodes sampled from $W$. Then for $N$ large enough, with probability at least $1-\nu$:
$$
\Vert W-W_G\Vert_{L^2}\leq\sqrt[4]{2N}\sqrt{\phi(N)} .
$$
\end{lemma}
\smallskip
\begin{proof}
First, we have
$$
\Vert W-W_G\Vert_\square\leq \vertiii{ T_W-T_{W_G}} \leq \phi(N) .
$$
It is easy to see that $W_G$ is a %random-free 
step graphon that takes values in $\{0,1\}$ and we can apply the inequality derived in \cite[Remark 10.8]{janson2013graphons}, so that: 
$$
\Vert W-W_G\Vert_{L^1}\leq\sqrt{2N}\Vert W-W_G\Vert_\square\leq\sqrt{2N}\phi(N) .
$$
Applying the relation $\Vert W-W_G\Vert_{L^2}\leq\Vert W-W_G \Vert_{L^1}^{1/2}$ yields the desired result.
%Then, we obtain:
%$$
%\Vert W-W_G\Vert_{L^2}^2\leq\Vert W-W_G\Vert_{L^1}\leq\sqrt{2N}\phi(N) .
%$$
%Finally, we get:
%$$
%\Vert W-W_G\Vert_{L^2}\leq\sqrt[4]{2N}\sqrt{\phi(N)} .
%$$
\end{proof}

\section{SIS Epidemics}

We consider a deterministic SIS model over a network modeled by a connected graph $G$ with homogeneous recovery and infection rates. The dynamics of each agent can be modeled by \cite{nowzari2016analysis}: 
$$
\dot{x}_i(t)=-\delta x_i(t)+\sum_{j=1}^Na_{ij}\beta x_j(t)(1-x_i(t)) ,
$$  
where $x_i(t)\in[0,1]$ is the fraction of the $i$th subpopulation that is infected at time $t$, $\delta$ is the recovery rate and $\beta$ is the infection rate.  
The model of all the network can be expressed in vector form as:
\begin{equation}\label{SISnonlinear}
\dot{x}(t)=(\beta A-\delta I)x(t)-\beta X(t)Ax(t) ,
\end{equation}
where $x(t)=[x_1(t),x_2(t),\ldots,x_N(t)]^T$ is the state vector of the system, $I$ is the identity matrix, $X(t)=\mathrm{diag}[x_1(t),x_2(t),\dots,x_N(t)]$ is a diagonal matrix and $A$ is the adjacency matrix of the network.

\subsection{Stability of SIS epidemics}
For any initial condition, the equilibrium $x=0$ (disease-free state) is globally asymptotically stable if 
%%%and only if 
%\margin{It  would be nice to write IF AND ONLY IF instead of IF.
%The cited theorem is IF AND ONLY IF, but has $\le$ instead of $<$.
%The problem is that the linearized case, instead, has $<$.
%So, to avoid long explanations, we only state the IF part, with $<$,
%which is true both for nonlinear and for linearized}
the following condition is satisfied \cite[Theorem 6]{nowzari2016analysis}: 
\begin{equation}
\label{threshold}
\lambda_1(A)\dfrac{\beta}{\delta}<1 .
\end{equation} 

\smallskip
In the perspective approximating graphs with graphons, it is convenient to consider sequences of graphs parametrized by their size $N$.
Therefore, it is reasonable to assume that parameters $\delta$ or $\beta$ be also dependent on $N$: upon need, we shall emphasize this dependence by writing $\delta_N$ and $\beta_N$.

\smallskip
As a first example of application of graphons, we determine a condition to reach the disease-free state for a graph sampled from $W$, based on the characteristics of the graphon.
\begin{prop}[Stability of epidemics]\label{propo1}
Let $W$ be a piecewise Lipschitz graphon and $G$ a graph with $N$ nodes sampled from $W$, representing the network for the SIS epidemic modeled in \eqref{SISnonlinear}. Then for $N$ large enough, the epidemic will reach the disease-free state with probability at least $1-\nu$ if:
\begin{equation}\label{eq_estabilidad}
\delta_N>N \beta_N (\vertiii{T_W}+\phi(N)) .
\end{equation}
\end{prop}
\smallskip
\begin{proof}
Condition \eqref{threshold} is equivalent to
$
\delta>\beta\lambda_1(A).$ 
By Lemma \ref{lemma1}, with $N$ large enough and probability at least $1-\nu$:
$$
\left\vert \lambda_1\left(T_{W_G}\right)- \lambda_1(T_W) \right\vert \leq \phi(N) ,
$$ 
$$
\lambda_1\left(T_{W_G}\right)\leq \lambda_1(T_W)+\phi(N) .
$$
Since $\lambda_1(A)=N\lambda_1\left(T_{W_G}\right)$, the above inequality yields \eqref{eq_estabilidad}.
\end{proof}

\subsection{An Index for the Sensitivity to Noise}
When the focus is stability of the disease-free state, it is natural to study the linearization of \eqref{SISnonlinear} near the origin \cite{khanafer2014optimal}:
\begin{equation}
\label{SISmodel}
\dot{x}(t)=(\beta A-\delta I)x(t)=\mathcal{A}x(t).
\end{equation}
This linearization is exponentially stable \cite{mei2017dynamics} under condition~\eqref{threshold}. We now turn our attention to the robustness of this stability property and more precisely to quantifying how the epidemics react to noise in the neighborhood of the equilibrium. Indeed, noise can represent migrations or other phenomena that are not included in the original model \cite{forgoston2011maximal}.  
%Along the lines of multiple works both on epidemics \cite{forgoston2011maximal,pare2017epidemic} and on general network systems \cite{garin2010survey,fagnani2018introduction}, 
To this purpose, we include additive noise in \eqref{SISmodel} and define
\begin{equation}
\label{eq_noise}
\dot{x}(t)=\mathcal{A}x(t)+n(t),
\end{equation}
where $n(t)\in\mathbb{R}^N$ is a stochastic noise process. 
A measure of this sensitivity can be defined as the asymptotic mean-square error:
\begin{equation}\label{eq_index}
J^{\mathrm{noise}}_G:=\lim_{t\to\infty} \dfrac{1}{N}\mathbb{E}\left[\|x(t)\|_2^2\right] .
\end{equation}
Under suitable assumptions on the noise vector\footnote{Under the assumptions of Proposition~\ref{prop:noisy-index}, noisy system~\eqref{eq_noise} can have negative states, which lack physical meaning: in this case, system~\eqref{eq_noise} should be interpreted as a purely mathematical construct whose purpose is quantifying the sensitivity of system~\eqref{SISnonlinear} to small perturbations in a neighborhood of the disease-free state. % \footnote{During the review process, it has been brought to our attention that the recent paper \cite{pare2017epidemic} defines a state-dependent noise model that avoids negative states. We note that such a noise model preserves asymptotic stability of the system, thus making index~\eqref{eq_index} trivially zero and therefore uninformative.}%
However, a different choice of the noise model can avoid negative states. One option is the state-dependent noise defined in \cite{pare2017epidemic}: this noise model preserves asymptotic stability of the system, thus making index~\eqref{eq_index} trivially zero and therefore uninformative. Another option is taking an uncorrelated positive noise: this choice entails a positive mean $m$ and yields $$J^{\mathrm{noise}}_G\le \dfrac{\sigma^2}{2N}\sum_{i=1}^N\dfrac{1}{\delta-\beta\lambda_i(A)} + \dfrac{m^2}{(\delta-\beta\lambda_1(A))^2}.$$ 
The analysis of this upper bound follows the same considerations that we have developed for \eqref{condthreshold}.}, 
%The derivation and analysis of this upper bound is analogous to \eqref{condthreshold}.}, 
this noise index is determined by the eigenvalues of the network.

\begin{prop}\label{prop:noisy-index}
Consider the system given in \eqref{eq_noise} satisfying condition \eqref{threshold} and a noise vector with zero mean and autocorrelation function $\mathbb{E}[n(t)n(t-\xi)^T]=\sigma^2\deltaup(\xi)I$. Then, the noise index~\eqref{eq_index} can be expressed as:
\begin{equation}\label{condthreshold}
J^{\mathrm{noise}}_G=\dfrac{\sigma^2}{2N}\sum_{i=1}^N\dfrac{1}{\delta-\beta\lambda_i(A)} .
\end{equation}
\end{prop}

%\mbox{}
\smallskip
\begin{proof}
% The solution of system \eqref{eq_noise} is given by:
% $$
% x(t)=e^{\mathcal{A}t}x(0)+\int_0^t e^{\mathcal{A}(t-\tau)}n(\tau)d\tau .
% $$
Considering that the solution of system \eqref{eq_noise} is 
$
x(t)=e^{\mathcal{A}t}x(0)+\int_0^t e^{\mathcal{A}(t-\tau)}n(\tau)d\tau,$ 
we calculate the expected value of $\Vert x(t)\Vert_2^2$:
\blue{
\begin{align}
 \nonumber
\mathbb{E}\left[\Vert x(t)\Vert_2^2\right]=&
 \mathbb{E}\left[ \left\Vert e^{\mathcal{A}t}x(0)+\int_0^t e^{\mathcal{A}(t-\tau)}n(\tau)d\tau\right\Vert_2^2\right]\\
\label{eq:three-terms}
=&\mathbb{E}\left[\left\Vert e^{\mathcal{A}t}x(0)\right\Vert_2^2\right]\\
\nonumber
 &+2\, \mathbb{E}\left[\left(e^{\mathcal{A}t}x(0)\right)^T\int_0^t e^{\mathcal{A}(t-\tau)}n(\tau)d\tau\right]\\
 \nonumber  &+\mathbb{E}\left[ \left(\int_0^t e^{\mathcal{A}(t-\tau)}n(\tau)d\tau\!\right)^{\!T} \int_0^t e^{\mathcal{A}(t-\tau)}n(\tau)d\tau\right].
 %
 %
%\mathbb{E}\left[\Vert x(t)\Vert_2^2\right]&=\mathbb{E}\left[ \left\Vert e^{\mathcal{A}t}x(0)+\int_0^t e^{\mathcal{A}(t-\tau)}n(\tau)d\tau\right\Vert_2^2\right]\\
%&=\mathbb{E}\left[\left\Vert e^{\mathcal{A}t}x(0)\right\Vert_2^2+2\left(e^{\mathcal{A}t}x(0)\right)^T\int_0^t e^{\mathcal{A}(t-\tau)}n(\tau)d\tau\right.\\
%&\qquad + \left. \left(\int_0^t e^{\mathcal{A}(t-\tau)}n(\tau)d\tau\right)^T \left(\int_0^t e^{\mathcal{A}(t-\tau)}n(\tau)d\tau\right)\right] .
\end{align}
}% end of blue
We begin by studying the third term (hereby denoted by $U$):
\begin{align*}
U
% &:=\mathbb{E}\left[ \left(\int_0^t e^{\mathcal{A}(t-\tau)}n(\tau)d\tau\right)^T \left(\int_0^t e^{\mathcal{A}(t-\tau)}n(\tau)d\tau\right)\right] \\
% %&=\mathbb{E}\left[ \int_0^t \int_0^t\left(e^{\mathcal{A}(t-\tau_1)}n(\tau_1)\right)^T\left(e^{\mathcal{A}(t-\tau_2)}n(\tau_2)\right)d\tau_1 d\tau_2\right]\\
&=\int_0^t \int_0^t \mathbb{E}\left[\left(e^{\mathcal{A}(t-\tau_1)}n(\tau_1)\right)^T\left(e^{\mathcal{A}(t-\tau_2)}
n(\tau_2)\right)\right]d\tau_1 d\tau_2 .
\end{align*}
Since for a real scalar $a$, we have $a=\mathrm{tr}(a)$, we can apply the trace to the integrand and its cyclic property, obtaining:
\begin{align*}
U&=\int_0^t \int_0^t \mathbb{E}\left[\mathrm{tr}\left(e^{\mathcal{A}(t-\tau_1)} e^{\mathcal{A}(t-\tau_2)}n(\tau_2)n(\tau_1)^T\right)\right]d\tau_1 d\tau_2\\
&=\int_0^t \int_0^t \mathrm{tr}\left(e^{\mathcal{A}(t-\tau_1)} e^{\mathcal{A}(t-\tau_2)}\mathbb{E}\left[n(\tau_2)n(\tau_1)^T\right]\right)d\tau_1 d\tau_2 .
\end{align*}
Due to the characteristics of the autocorrelation function of the noise, we have $\mathbb{E}\left[n(\tau_2)n(\tau_1)^T\right]=\sigma^2\blue{\deltaup}(\tau_2-\tau_1)I$. Thus:     
\begin{align*}
U&=\int_0^t \int_0^t \mathrm{tr}\left(e^{\mathcal{A}(t-\tau_1)} e^{\mathcal{A}(t-\tau_2)}\sigma^2\blue{\deltaup}(\tau_2-\tau_1)I\right)d\tau_1 d\tau_2\\
&=\sigma^2\mathrm{tr}\left( \int_0^t \left(\int_0^t  \blue{\deltaup}(\tau_2-\tau_1)e^{\mathcal{A}(t-\tau_1)} d\tau_1\right) e^{\mathcal{A}(t-\tau_2)}d\tau_2\right) .
\end{align*}
%----- old
% Using the property of the integral of Dirac's delta function $\int_{-\infty}^\infty f(x) \blue{\deltaup}(x-a)dx=f(a)$, we get:
% \begin{align*}
% U&=\sigma^2\mathrm{tr}\left( \int_0^t \left(e^{\mathcal{A}(t-\tau_2)}\right) e^{\mathcal{A}(t-\tau_2)}d\tau_2\right)\\
% &=\sigma^2\int_0^t \mathrm{tr}\left(e^{2\mathcal{A}(t-\tau_2)} \right)d\tau_2 .
% \end{align*}
%--------
%------- new, shorter
Using linearity of the trace and \blue{Dirac's delta property that $\int_{I_a} \deltaup(\tau-a) \varphi(\tau) d \tau = \varphi(a)$ for any interval $I_a$ that has $a$ in its interior and any test function $\varphi$}, %$\int_{-\infty}^\infty f(x) \blue{\deltaup}(x-a)dx=f(a)$
we get:
\[
U =\sigma^2\int_0^t \mathrm{tr}\left(e^{2\mathcal{A}(t-\tau_2)} \right)d\tau_2 .
\]
%--------
Being matrix $\mathcal{A}$ symmetric, it can be written $\mathcal{A}=Q\Lambda Q^T$ where $Q$ is an orthogonal matrix and $\Lambda$ is a diagonal matrix with the eigenvalues of $\mathcal{A}$. Since $e^{\mathcal{A}}=Qe^{\Lambda}Q^T$, we have:
\begin{align*}
U&=\sigma^2\int_0^t \mathrm{tr}\left(Qe^{2\Lambda(t-\tau_2)}Q^T \right)d\tau_2\\
&=\sigma^2\int_0^t \mathrm{tr}\left(e^{2\Lambda(t-\tau_2)} \right)d\tau_2\\
&=\sigma^2\int_0^t \sum_{i=1}^N e^{2(t-\tau_2)\lambda_i(\mathcal{A})}d\tau_2\\
&=\sigma^2\sum_{i=1}^N e^{2t\lambda_i(\mathcal{A})}\int_0^t e^{-2\tau_2\lambda_i(\mathcal{A})}d\tau_2\\
&=\sigma^2\sum_{i=1}^N \dfrac{1}{2\lambda_i(\mathcal{A})}\left[e^{2t\lambda_i(\mathcal{A})}-1\right] .
\end{align*}
\blue{We observe that the first term in \eqref{eq:three-terms} becomes zero when $t\rightarrow \infty$ because of condition \eqref{threshold}, while the second term is zero because the expected value of the noise is zero.
Then,}
$$
J^{\mathrm{noise}}_G=\dfrac{1}{N}\lim_{t\to\infty} \left(\sigma^2\sum_{i=1}^N \dfrac{1}{2\lambda_i(\mathcal{A})}\left[e^{2t\lambda_i(\mathcal{A})}-1\right]\right).
$$
\blue{The conclusion follows because condition \eqref{threshold} implies that all eigenvalues of $\mathcal{A}$ are negative. }
% Due to condition \eqref{threshold}, all the eigenvalues of matrix $\mathcal{A}$ are negative and \blue{therefore
% $
% J^{\mathrm{noise}}_G=-\dfrac{\sigma^2}{2N}\sum\limits_{i=1}^N\dfrac{1}{\lambda_i(\mathcal{A})},$ as desired.} %=\dfrac{\sigma^2}{2N}\sum_{i=1}^N\dfrac{1}{\delta-\beta\lambda_i(A)} .$$
\end{proof}

% %-----------old
% In this context, the objective is to estimate this noise index for a graph sampled from a graphon using the spectrum of the graphon operator. If we consider a graphon with finite rank $M$, it is possible to estimate the noise index $J^{\mathrm{noise}}_G$ of a network $G$ with $N$ nodes sampled from $W$, by defining a similar quantity as:
% $$
% J^{\mathrm{noise}}_{W,N}:=\dfrac{\sigma^2}{2N}\sum_{i=1}^N \dfrac{1}{\delta-N\beta\lambda_i(T_W)} ,
% $$
% where $\lambda_i(T_W)$ are the nonzero eigenvalues of $T_W$ for $i=1,\ldots,M$ and $\lambda_i(T_W)=0$ for $i=M+1,\ldots,N$.
% %---------- end old

%\margin{check this discussion on $\beta_N/\delta_N$}
%------------new, including discussion on delta_N and beta_N
\smallskip
Our objective is to estimate this noise index for a graph sampled from a graphon, by using the spectrum of the graphon operator. Assuming that the graphon $W$ has finite rank $M$,
for all $N \ge M$ we define
 $$
 J^{\mathrm{noise}}_{W,N}:=\dfrac{\sigma^2}{2N}\sum_{i=1}^N \dfrac{1}{\delta_N-\beta_N N \lambda_i(T_W)} ,
 $$
 where $\lambda_i(T_W)$ are the nonzero eigenvalues of $T_W$ for $i=1,\ldots,M$ and $\lambda_i(T_W)=0$ for $i=M+1,\ldots,N$.
%   \margin{Should we cite (5) or (6)? 
% In Thm1 we assume (6), to ensure (5) with prob 1-nu.
% If we put (6) here, then below we should 
% say $\delta_N - N \beta_N (\vertiii{T_W}+\phi(N))$
% bounded away from 0,
% instead of $\lambda_1(A_N) \beta_N / \delta_N$ bounded away from 1.} %end margin
 We will use $J^{\mathrm{noise}}_{W,N}$ as an approximation of $J^{\mathrm{noise}}_G$
 when $G$ is a large graph sampled from $W$; Theorem~\ref{main_theorem} will ensure that the approximation error %tends to zero
 is small
 with high probability.
 Since this result involves epidemics on graphs of increasing size $N$, we have to specify the dependence on $N$ of the  parameters $\delta_N$ and $\beta_N$. \blue{Clearly, we need to satisfy the stability condition \eqref{threshold} for all $N$: actually, we will need the stronger property that $\lambda_1(A_N) \beta_N / \delta_N $ remains bounded away from $1$ also in the limit for $N \to \infty$. 
%-------------
% OLD
% We should bear in mind that graphons are approximations of {\em dense} graphs, that is, graphs whose degrees grow linearly with $N$.
% NEW
Since almost surely $\lambda_1(A_N)$  grows linearly with $N$ (except for the trivial graphon $W=0$), in order to}
%------------
 ensure a uniform bound on $\lambda_1(A_N) \beta_N / \delta_N $, we will assume
that the infection strength $\beta_N/\delta_N$ satisfies
\begin{equation} \label{eq_betaN_deltaN}
    \frac{\beta_N}{\delta_N}=\frac1N \frac{\bar \beta}{\bar \delta}
\end{equation}
for some positive constants $\bar \beta$ and $\bar \delta$.
%---------------

% %--------------- old discussion on scaling, by Paolo
% TO BE MOVED 
% When considering sequences of graphs of growing size, we should bear in mind that graphons are approximations of {\em dense} graphs, that is, graphs whose nodes' degrees grow linearly with $N$. Under this assumption, $\lambda_1(A)$ will be linearly increasing with $N$ and therefore in order to satisfy \eqref{threshold} one could assume a relation like 
% $$\frac{\beta_N}{\delta_N}=\frac1N \frac{\bar \beta}{\bar \delta},$$
% where  $\bar \delta$ and $\bar \beta$ are positive constants (independent of $N$).
% %-------------

\begin{theorem}[Graphon approximation]\label{main_theorem}
Let $W$ be a piecewise Lipschitz graphon with finite rank $M$ and $G$ a graph with $N$ nodes sampled from $W$ with $N \geq M$. If \blue{$\delta_N=\bar{\delta}$ and $\beta_N=N^{-1}\bar{\beta}$} satisfy condition \eqref{eq_estabilidad}, then for $N$ large enough, with probability at least $1-\nu$:
\begin{align*}
\Delta^{\mathrm{noise}}&:=\left\vert J^{\mathrm{noise}}_G-J^{\mathrm{noise}}_{W,N} \right\vert\\
&\leq\dfrac{1}{N^{\blue{3}/4}}\dfrac{\sigma^2\bar{\beta}\sqrt{\sqrt{N\log(2N/\nu)}+\sqrt{L^2-K^2+KN}}}{\sqrt[4]{2}(\bar{\delta}-\bar{\beta}\vertiii{T_{W}}-\bar{\beta}\phi(N))(\bar{\delta}-\bar{\beta}\vertiii{T_W})} .
\end{align*} 
\end{theorem}

\mbox{}

%To prove the theorem, we first need to obtain a bound in the $L^2$ norm for the difference between $W$ and $W_G$. 

\begin{proof}
\blue{Throughout this proof, we will assume that $\lambda_i(T_W)$ (i.e., the $M$ non-zero eigenvalues of $T_W$ and $N-M$ zeros) are sorted in non-increasing order: this choice does not change the sum that defines $J^{\mathrm{noise}}_{W,N}$ and is consistent with the non-increasing ordering of $\lambda_i(A)$. This consistency will be useful in order to apply Wielandt-Hoffman Theorem.}
%By definition:
% $$
% \Delta^{\mathrm{noise}}=\left\vert\dfrac{\sigma^2}{2N} \sum_{i=1}^N \dfrac{1}{\delta_N-\beta_N\lambda_i(A)}-\dfrac{\sigma^2}{2N}\sum_{i=1}^N \dfrac{1}{\delta_N-N\beta_N\lambda_i(T_W)}\right\vert .
% $$
Since $\lambda_i(A)=N\lambda_i\left(T_{W_G}\right)$, \blue{$\delta_N=\bar{\delta}$
and $\beta_N = N^{-1}\bar \beta$}, we have: 
\begin{align*}
\Delta^{\mathrm{noise}}
&=\left\vert\dfrac{\sigma^2}{2N} \sum_{i=1}^N \dfrac{1}{\delta_N-\beta_N\lambda_i(A)}-\dfrac{\sigma^2}{2N}\sum_{i=1}^N \dfrac{1}{\delta_N-N\beta_N\lambda_i(T_W)}\right\vert\\
&=\dfrac{\sigma^2}{2N}\left\vert \sum_{i=1}^N \dfrac{1}{\bar{\delta}-\bar \beta\lambda_i\left(T_{W_G}\right)}-\sum_{i=1}^N \dfrac{1}{\bar{\delta}-\bar \beta\lambda_i(T_W)}\right\vert\\ 
&=\dfrac{\sigma^2\bar{\beta}}{2N}\left\vert \sum_{i=1}^N \dfrac{\lambda_i(T_W)-\lambda_i\left(T_{W_G}\right)}{\left(\bar{\delta}-\bar \beta\lambda_i\left(T_{W_G}\right)\right)(\bar{\delta}-\bar \beta\lambda_i(T_W))}\right\vert\\
&\leq\dfrac{\sigma^2\bar{\beta}}{2N} \sum_{i=1}^N \dfrac{\left\vert\lambda_i(T_W)-\lambda_i\left(T_{W_G}\right)\right\vert}{\left(\bar{\delta}-\bar \beta\lambda_i\left(T_{W_G}\right)\right)(\bar{\delta}-\bar \beta\lambda_i(T_W))}\\
&\leq\dfrac{\sigma^2\bar{\beta} \sum_{i=1}^N \left\vert\lambda_i(T_W)-\lambda_i\left(T_{W_G}\right)\right\vert}{2N\left(\bar{\delta}-\bar \beta\vertiii{T_{W_G}}\right)(\bar{\delta}-\bar \beta\vertiii{T_W})} . 
\end{align*}
We define the vector $\lambda_{T_W-T_{W_G}}$ as:
$$
\lambda_{T_W-T_{W_G}}=\left[\lambda_1(T_W)-\lambda_1\left(T_{W_G}\right), \ldots, \lambda_N(T_W)-\lambda_N\left(T_{W_G}\right)\right]^T  ,
$$
so that the sum in the numerator is 
$$
\sum_{i=1}^N \left\vert\lambda_i(T_W)-\lambda_i\left(T_{W_G}\right)\right\vert=\left\Vert \lambda_{T_W-T_{W_G}} \right\Vert_1 .
$$
Then, \blue{the relation %between vector norms
$\Vert \cdot \Vert_1 \leq \sqrt{N}\Vert \cdot \Vert_2$ implies}
$$
\Delta^{\mathrm{noise}}\leq\dfrac{\sigma^2\bar{\beta} \left(\sum_{i=1}^N \left\vert\lambda_i(T_W)-\lambda_i\left(T_{W_G}\right)\right\vert^2\right)^{1/2} }{2N^{1/2}\left(\bar{\delta}-\bar \beta\vertiii{T_{W_G}}\right)(\bar{\delta}-\bar \beta\vertiii{T_W})} .
$$
We can apply Wielandt-Hoffman Theorem in infinite dimensional spaces \cite[Theorem 2]{bhatia1994hoffman} and, \blue{since $N\ge M$,} get:
$$
\Delta^{\mathrm{noise}}\leq\dfrac{\sigma^2\bar{\beta} \left\Vert T_W-T_{W_G} \right\Vert_{\mathrm{HS}}}{2N^{1/2}\left(\bar{\delta}-\bar \beta\vertiii{T_{W_G}}\right)(\bar{\delta}-\bar \beta\vertiii{T_W})} .
$$
Since $\left\Vert T_W-T_{W_G} \right\Vert_{\mathrm{HS}}=\Vert W-W_G \Vert_{L^2}$, \blue{Lemma~\ref{lem_bnorm} implies that}
$$
\Delta^{\mathrm{noise}}\leq\dfrac{\sigma^2\bar{\beta}\sqrt[4]{2N}\sqrt{\phi(N)}}{2N^{1/2}\left(\bar{\delta}-\bar \beta\vertiii{T_{W_G}}\right)(\bar{\delta}-\bar \beta\vertiii{T_W})},
$$
%\begin{align*}
%\Delta^{\mathrm{noise}}&\leq\dfrac{\sigma^2\sqrt[4]{2N}\sqrt{\phi(N)}}{2N^{3/2}\left(\bar{\delta}-\beta\vertiii{T_{W_G}}\right)(\bar{\delta}-\beta\vertiii{T_W})} \\
%&\leq\dfrac{1}{N^{7/4}}\dfrac{\sigma^2\sqrt{\sqrt{N\log(2N/\nu)}+\sqrt{L^2-K^2+KN}}}{\sqrt[4]{2}\left(\bar{\delta}-\beta\vertiii{T_{W_G}}\right)(\bar{\delta}-\beta\vertiii{T_W})} .
%\end{align*}
\blue{and the proof is completed by using the definition of $\phi(N)$ and applying Lemma~\ref{lemma1} in the denominator.}
%$$
%\Delta^{\mathrm{noise}}\leq\dfrac{1}{N^{7/4}}\dfrac{\sigma^2\sqrt{\sqrt{N\log(2N/\nu)}+\sqrt{L^2-K^2+KN}}}{\sqrt[4]{2}(\bar{\delta}-\beta\vertiii{T_{W}}-\beta\phi(N))(\bar{\delta}-\beta\vertiii{T_W})} .
%$$
\end{proof}

\blue{Assuming that $\delta=\bar \delta$ is constant and $\beta=N^{-1}\bar\beta$ means that, as the graph grows in size, the healing rate (which depends on each individual) remains constant, whereas the infection rate decreases. This natural scaling law is also chosen in \cite{gao2019spectral}. Indeed, on dense graphs this assumption means that in  larger graphs, even though there are more potential interactions, the average strength of the connections is suitably adjusted: this fact accounts for natural limitations in the rates of contact between individuals.}

\begin{remark}[Asymptotics for $N\to\infty$ \& scaling factors]\label{remark1}
\blue{
When  we let $N$ go to infinity, if $\nu$ is constant or if $\nu = N^\alpha$ for some constant $\alpha$,
we can see that the upper bound given in Theorem~\ref{main_theorem} for
the estimation error $\Delta^{\mathrm{noise}}$
goes to zero as $O\left((\log N)^{1/4}/(N^{1/2})\right)$.
Hence, by choosing $\alpha > 1$ and applying Borel-Cantelli Lemma, we obtain that $\Delta^{\mathrm{noise}}$
 almost surely converges to zero with rate $O\left((\log N)^{1/4}/(N^{1/2})\right)$.
Index $J^{\mathrm{noise}}_G$ is bounded and bounded away from zero under the assumptions of Theorem~\ref{main_theorem}.
Hence, the same asymptotic behaviour holds for
the relative error $\Delta^{\mathrm{noise}}/J^{\mathrm{noise}}_G$.}
This asymptotic behaviour of the relative error
does not depend on the assumption \blue{$\beta_N = N^{-1} \bar \beta$}
and remains true for any choice of $\delta_N$ and $\beta_N$
that satisfies  \eqref{eq_estabilidad} and \eqref{eq_betaN_deltaN}. 
Indeed, any other choice of $\delta_N$ and $\beta_N$ 
satisfying \eqref{eq_betaN_deltaN} would modify
both $J^{\mathrm{noise}}_G$ and $J^{\mathrm{noise}}_{W,N}$
%(and hence also $\Delta^{\mathrm{noise}}$)
by the same multiplicative factor,
so that the relative error would remain the same.
\end{remark}

\section{Numerical and Simulation Results}
\begin{figure}%[!b]
\centering
\includegraphics[width=3.3cm]{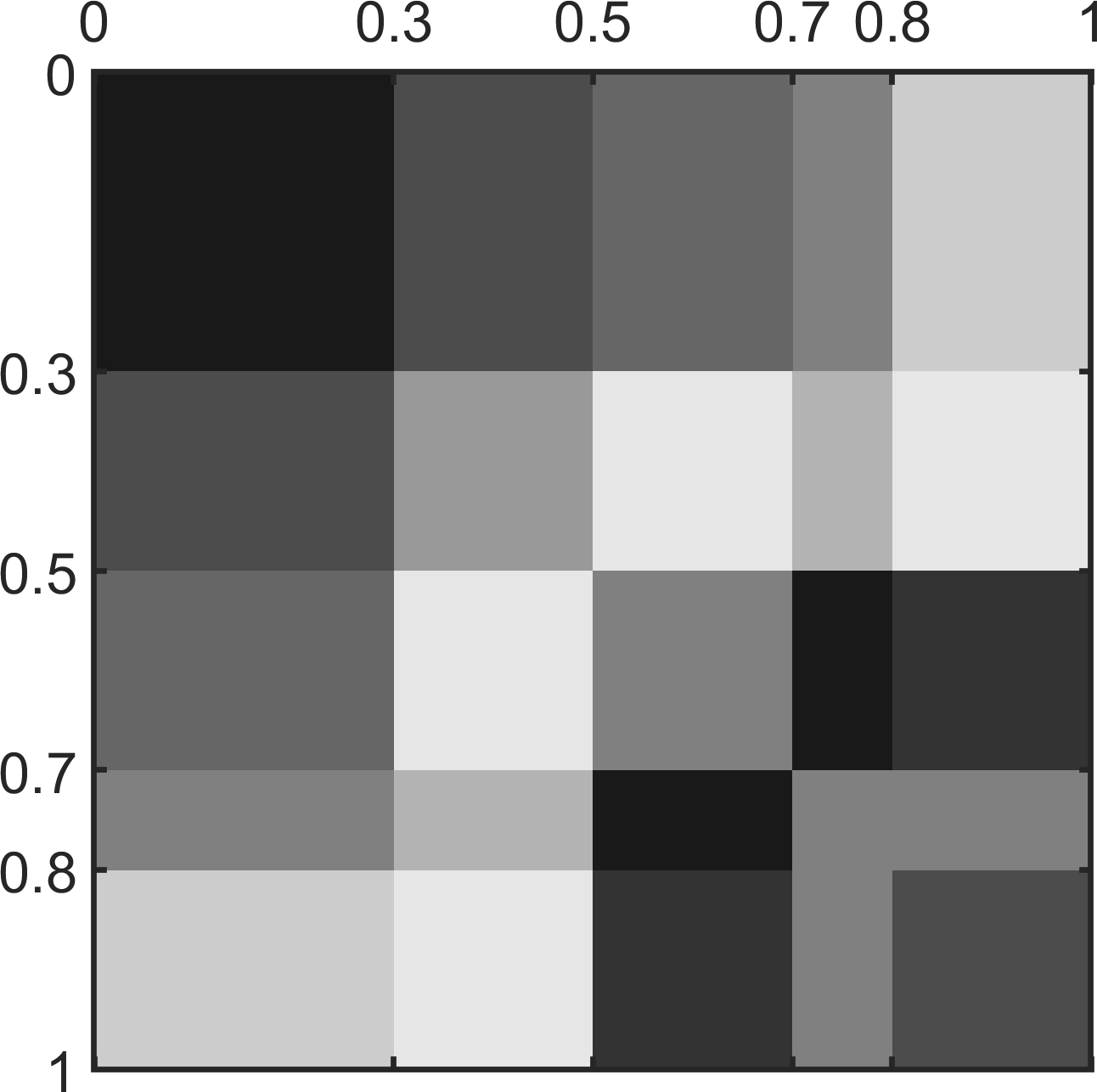} 
\caption{Pixel diagram of the stochastic block model graphon.}
\label{block_graphon}
\end{figure}

\begin{figure}%[!h]
\centering
\includegraphics[width=\columnwidth]{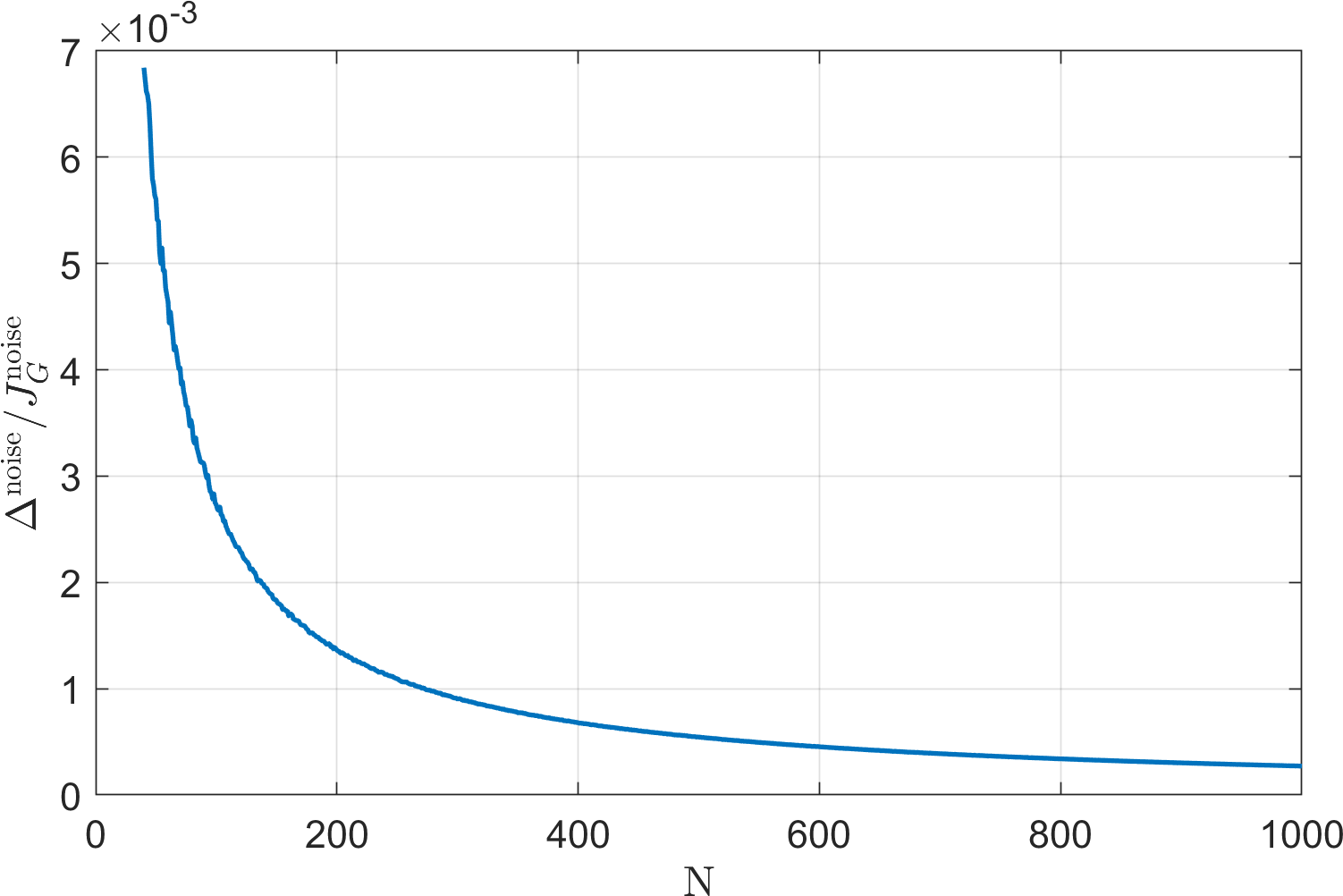}
\caption{Relative error $\Delta^{\mathrm{noise}}/J^{\mathrm{noise}}_G$ as a function of $N$ for the stochastic block model in Fig.~\ref{block_graphon}.}
\label{percent}
\end{figure}

We consider the stochastic block model graphon $W_{\mathrm{SB}}$ with pixel diagram in Fig.~\ref{block_graphon}, where the values of the blocks are:
$$
W_{\mathrm{SB}}=
\begin{bmatrix}
   0.9 & 0.7 & 0.6 & 0.5 & 0.2 \\
   0.7 & 0.4 & 0.1 & 0.3 & 0.1 \\
   0.6 & 0.1 & 0.5 & 0.9 & 0.8 \\
   0.5 & 0.3 & 0.9 & 0.5 & 0.5 \\
   0.2 & 0.1 & 0.8 & 0.5 & 0.7  
\end{bmatrix}.
$$ 
The nonzero eigenvalues of the graphon operator $T_{W_{\mathrm{SB}}}$ are $\lambda_1=0.5275$, $\lambda_2=0.2098$, $\lambda_3=0.0126$, $\lambda_4=-0.0128$ and $\lambda_5=-0.0971$. We assume  $\beta_N=\bar\beta=0.1$ and $\nu=0.02$, which satisfies condition \eqref{eq_condi3} for $N\geq 40$, and we generate sampled graphs from $W_{SB}$ with $40\leq N \leq 1000$. 
%\margin{Think notation: should we use $\bar \beta$, $\delta_N$? Is $\delta_N = N \bar \delta$ for some $\bar \delta$?}
For each network, the value of $\delta_N=N\bar\delta=N\bar\beta(\vertiii {T_W}+\phi(40))$ is selected, satisfying condition \eqref{eq_estabilidad}, and we compute the relative error $\Delta^{\mathrm{noise}}/J^{\mathrm{noise}}_G$, obtaining the results of Fig.~\ref{percent}. As per Remark~\ref{remark1}, the relative error goes to zero as $N$ increases. 

%validating the rates of convergence of Remark \ref{remark1}.    

% Finally, we compare the value of the noise index for the linearized model an analogous noise index considering the nonlinear dynamics 
%  \begin{equation}
%  \label{eq_noise_nonlinear}
%  \dot{x}(t)=(\beta A-\delta I)x(t)-\beta X(t)Ax(t)+n(t).
%  \end{equation}
% To this purpose, we perform 100 realizations of the process for networks with $N=50$, 100, 200 and 500 nodes and we obtain the results of Table \ref{tab1}. We can see that the values are close and the error is negligible. The index is not zero for all the cases, implying a fraction of people who remains infected as a consequence of the noise. 

% \begin{table}%[!h]
% \caption{Noise index for linear and nonlinear models.}
% \begin{center}
% %\resizebox{.48\textwidth}{!}{
%  \begin{tabular}{|c|*{3}{c|}} 
%  \hline \multirow{2}{*}{$\bm{N}$}& \multicolumn{2}{c|}{\textbf{Noise index} }&\multirow{2}{*}{\textbf{Relative error}}\\
%  \cline{2-3} & \textbf{Linear} & \textbf{Nonlinear} & \\ 
%   \hline
%   50 & $2.630\times 10^{-5}$  & $2.654\times 10^{-5}$ & $0.00909$  \\ 
%   \hline
%   100  & $1.634\times 10^{-5}$  & $1.629\times 10^{-5}$ & $0.00275$ \\
%   \hline
%   250  & $8.669\times 10^{-6}$  & $8.699\times 10^{-6}$ & $0.00347$ \\
%   \hline
%   500  & $5.303\times 10^{-6}$  & $5.293\times 10^{-6}$ & $0.00188$ \\
%   \hline
% \end{tabular}
% \end{center}
% \label{tab1}
% \end{table}

\section{Conclusion}

% OLD: This work presented a study and application of the properties of graphons to the analysis of SIS epidemics on large networks. The study of an epidemic of SIS type over a network sampled from a graphon was performed using the spectrum of the graphon operator. A condition to reach the disease-free state was derived considering the operator norm of a piecewise Lipschitz graphon. A performance index related to the presence of additive noise in the epidemic was developed based on the eigenvalues of the operator associated with a piecewise Lipschitz graphon with finite rank. 
%
%During the development of the work, many questions \blue{arised.} One of the aspects that require further investigation is the extension of the results obtained in this work to networks sampled from graphons which do not have a finite rank: \blue{this extension could be enabled by the spectral approximation tools developed in \cite{gao2019spectral}}. Also, we would like to develop performance indices for epidemics with more compartments such as SIR models.

\blue{This work presented an analysis of SIS epidemics on large networks, under the assumption that the network is sampled from a graphon. Relevant information about the stability of an epidemic can be inferred from the graphon, without the need to perform computations on, or even know, the full network topology. In this vein, we have derived a stability criterion in Proposition~\ref{propo1} and defined a noise-sensitivity index (Theorem~\ref{main_theorem}) that both only depend on the graphon.

Several questions are left open by this work, including the extension of our results to networks sampled from graphons that do not have finite rank: this extension could be enabled by the spectral approximation tools developed in \cite{gao2019spectral}. 
Moreover, we believe that graphons can help not only the analysis but also the control of epidemics: for instance, graphon centrality~\cite{avella2018centrality} can provide guidance for targeted interventions such as quarantine or vaccination.}
%Also, we would like to develop performance indices for epidemics with more compartments such as SIR models.


%%%%%%%%%%%%%%%%%%%%%%%%%%%%%%%%%%%%%%%%%%%%%%%%%%%%%%%%%%%%%%%%%%%%%%%%%%%%%%%%

%%%%%%%%%%%%%%%%%%%%%%%%%%%%%%%%%%%%%%%%%%%%%%%%%%%%%%%%%%%%%%%%%%%%%%%%%%%%%%%%

%%%%%%%%%%%%%%%%%%%%%%%%%%%%%%%%%%%%%%%%%%%%%%%%%%%%%%%%%%%%%%%%%%%%%%%%%%%%%%%%

 %\section*{Acknowledgment}

 %This work has been partially supported by the LabEx PERSYVAL-Lab (ANR-11-LABX-0025-01) funded by the French program Investissement d'Avenir and by ANR grant HANDY (ANR-18-CE40-0010).

%\newcommand{\newblock}{}
\bibliographystyle{IEEEtran}
%\bibliography{references}

\end{document}